# Geometrical comparison of two protein structures using Wigner-D functions

S. M. Saberi Fathi[1] and Jack A. Tuszynski[2]

[1] Department of Physics, Ferdowsi University of Mashhad, Mashhad, Iran

Email: saberifathi@um.ac.ir

[2] Department of Physics, University of Alberta, Edmonton, Alberta, Canada

Email: jackt@ualberta.ca

**Abstract**

In this paper, we develop a quantitative comparison method for two arbitrary protein structures. This method uses a root-mean-square deviation (RMSD) characterization and employs a series expansion of the protein's shape function in terms of the Wigner-D functions to define a new criterion, which is called a "similarity value". We further demonstrate that the expansion coefficients for the shape function obtained with the help of the Wigner-D functions correspond to structure factors. Our method addresses the common problem of comparing two proteins with different numbers of atoms. We illustrate it with a worked example.

**Introduction**

A quantitative comparison of two protein tertiary structures to assess their similarity is a major challenge, but if properly investigated, it can offer answers to important questions in biochemistry and cell biology[1]. In particular, structural similarity between proteins is a very good predictor of their functional similarity. In order to classify proteins according to their structural characteristics, we first have to be able to determine the 3D structures of the proteins in question, which typically involves x-ray or electron crystallography, or in some cases other techniques such as nuclear magnetic resonance (NMR) or mass spectroscopy[2]. In the absence of crystallographic structures for a given protein, computational methods may still be used to predict a 3D structure based on sequence similarity with crystallographically resolved protein structures using a technique called homology modeling[3]. Assuming structural information is available, a number of methods have been developed to compare protein structures[4,5]. Some methods are based on numerical techniques such as geometric hashing[6] or spherical harmonic descriptors [7]. A recently reported method uses so-called Zernike descriptors[8].

Traditionally, protein classifications have been performed manually with the aid of automated tools, and they take into account information available to biologists regarding both the functions and the phylogenetic origins of the proteins investigated. Examples of relevant databases include

SCOP (Structural Classification of Proteins)[9,10], CATH (Class, Architecture, Topology, and Homologous superfamily)[11], and FSSP (Families of Structurally Similar Proteins)[12] to name but a few.

In order to match two distinct protein structures, there should exist a one-to-one map between their structural elements, which is called "correspondence". In addition, proper alignment of the structural elements of these proteins should be generated. A common measure that is used for this type of alignment is RMSD[13,14]. Until now, a complete geometrical comparison of two proteins has rarely been possible mainly because most proteins have different sizes and/or different numbers and types of atoms. Therefore, a complete match between an arbitrary pair of proteins is a difficult task to accomplish in general. This is why either partial or local similarity tests have frequently been used in the past[13]. An example of using RMSD for partial similarity analysis is the STRUCTAL software[15]. In the Results and Discussion section, we discuss in more detail different methods used for protein structure comparisons and compare and contrast them with our method

In this paper, we introduce a fully automated method that enables one to compare protein structures and to perform identification of proteins. To this end we expand the protein shape function in terms of Wigner-D functions[16] and demonstrate mathematically that the expansion coefficients can be regarded as the structure factors of a protein. We then compare them to assess their similarity by introducing a new parameter referred to here as the "Similarity Value" (*SV*). Our method obtains the similarity value in the reciprocal space (in relation to the spatial domain) where two proteins have the same dimension (values of their structure factors). However, it is important to note that these proteins are allowed to have different numbers of atoms in the spatial domain. We demonstrated below that the *SV* is generally a good alternative parameter to the RMSD value. However, in comparing different-size structures, using the similarity value (*SV*) is strongly preferred as it permits a quantitative comparison between any protein structures independently of their sizes.

**Basic mathematical idea**

The Wigner D-functions describe the rotation on a sphere in 4-dimensional space (4-sphere), and they are analogous to the well-known spherical harmonic functions, which are commonly used to describe the rotation on a sphere in 3-dimensional space (3-sphere)[16]. A rigid body can be projected on a 4-sphere; thus, its shape function can be expanded using the Wigner-D functions. Proteins are not typically thought of as rigid bodies due to their weak bonds, but instead they undergo sizeable thermal fluctuations at finite temperature and conformational changes due to ligand binding. However, the different conformations of a protein which are explored over time can be quantitatively characterized using shape functions in time series representations.

We start by expanding a hypothetical protein shape function, $f$, in terms of Wigner-D functions as

$$f(\alpha,\beta,\gamma) = \sum_{l=0}^{\infty} \sum_{m=-l}^{l} \sum_{n=-l}^{l} C_{lmn} D^l_{mn}(\alpha,\beta,\gamma)$$

(1)

where the $C_{lmn}$ factors are the coefficients of the series expansion, $D^l_{mn}$ is a Wigner-D function and the parameters $m, l$ and $n$ satisfy: $l \geq 0, |m| \leq l$ and $|n| \leq l$. The Wigner-D function is defined by[17]

$$D^l_{mn}(\alpha,\beta,\gamma) = e^{-im\alpha} e^{-im\gamma} d^l_{mn}(\cos\beta)$$

(2)

where

$$d^l_{mn}(x) = \varepsilon \sqrt{\frac{\left(l-\frac{\nu+\sigma}{2}\right)!\left(l+\frac{\nu+\sigma}{2}\right)!}{\left(l-\frac{\nu-\sigma}{2}\right)!\left(l+\frac{\nu-\sigma}{2}\right)!}} \; 2^{-\frac{\nu+\sigma}{2}} (1-x)^{\frac{\nu}{2}} (1+x)^{\frac{\sigma}{2}} P^{(\nu,\sigma)}_{l-\frac{\nu+\sigma}{2}}(x)$$

(3)

where $\nu = |n-m|$ and $\sigma = |n+m|$, and

$$\varepsilon = \begin{cases} 1 & \text{if } n \geq m \\ (-1)^{n-m} & \text{if } n < m \end{cases}$$

(4)

while $P^m_l(x)$ is the associated Legendre polynomial is defined as

$$P_l^m(x) = \frac{(-1)^m}{2^l \, l!} (1-x^2)^{\frac{m}{2}} \frac{d^{l+m}}{dx^{l+m}} (x^2-1)^l. \tag{5}$$

The dimension of a Wigner-D function is given by

$$Dim(D) = \sum_{l=0}^{N} (2l+1)^2 = \frac{1}{3}(N+1)(2N+1)(2N+3) \tag{6}$$

We can express Eq. 1 in matrix notation simply as $\mathbf{f} = \mathbf{CD}$. Indeed, the discrete Fourier transform on SO(3) can be written in terms of the Wigner-D functions as[17,18]

$$f(\alpha, \beta, \gamma) = \sum_{l=0}^{\infty} \sum_{m=-l}^{l} \sum_{n=-l}^{l} \hat{f}_{lmn} D^l{}_{mn}(\alpha, \beta, \gamma) \tag{7}$$

where $\hat{f}$ is the Fourier transform of $f$. We can express the above relation in matrix form as $\mathbf{f} = \mathbf{\hat{f}D}$. Thus, the $C_{lmn}$ coefficients can be viewed as Fourier transforms of a given function $f$. On the other hand, we know from crystallography that the Fourier transform of the shape function of an object is defined as the corresponding structure factor[19]. Thus, the $C_{lmn}$ coefficients describe the structure factors of a given protein with the shape function $f$ (which is obtained from the positions of the atoms of the protein).

Having generated the shape function $f$, we can obtain the $C_{lmn}$ coefficients of the expansion by

$$C_{lmn} = \frac{(2l+1)}{8\pi^2} \int \int \int f(\alpha, \beta, \gamma) \, D^l{}_{mn}{}^*(\alpha, \beta, \gamma) \, \sin\beta \, d\beta \, d\alpha \, d\gamma \tag{8}$$

where we use the orthogonality of the Wigner-D function:

$$\int \int \int D^{l'}{}_{m'n'}{}^*(\alpha, \beta, \gamma) \, D^l{}_{mn}(\alpha, \beta, \gamma) \, \sin\beta \, d\beta \, d\alpha \, d\gamma = \frac{8\pi^2}{(2l+1)} \delta_{ll'}\delta_{mm'}\delta_{nn'} \tag{9}$$

**Method and algorithm**

In this section we discuss practical aspects of implementing our method for particular proteins. First, we download each protein's atom positions from the Protein Data Bank (PDB) and convert these positions to obtain the corresponding Euler angles. Then, we define the protein shape function $f$ as follows: if a voxel contains a protein's atom then $f$ is equal to one, otherwise $f$ is taken to be zero. The next steps are to compute the resultant Wigner-D functions up to $l = l_{max}$ and obtain the $C_{lmn}$ matrix elements.

One simple way to measure the similarity between two arbitrarily selected proteins is equivalent to computing the correlation value between the structure factors of the two proteins:

$$\text{Corrlation Value} = CV = \frac{\langle \text{abs}(\mathbf{C}) | \text{abs}(\mathbf{C}'') \rangle}{\langle \text{abs}(\mathbf{C}) | \text{abs}(\mathbf{C}) \rangle \langle \text{abs}(\mathbf{C}') | \text{abs}(\mathbf{C}') \rangle}$$

(10)

where $\langle \cdot | \cdot \rangle$ indicates the inner product and $\text{abs}(\cdot)$ indicates the absolute value of a variable. However, the *CV* measure does not provide proper comparison results for proteins, as is explained below.

Representing a 3D shape by expansion in terms of Wigner-D functions effectively projects this shape on a 3-manifold as a part of the hyper-surface of a 4-sphere. The $C_{lmn}$ matrix elements are the points on the manifold constructed in this manner. The *CV* computed in Eq. 10 gives a fractional rate of the overlap between the two manifolds.

We illustrate this with a specific example. We have chosen a crystal structure for the tubulin heterodimer with PDB code 1JFF[20]. This PDB has two subunits: 1JFF-A for the α-tubulin monomer and 1JFF-B for the β-tubulin monomer. As shown in Table I, the *CV* for 1JFF and 1JFF-A is approximately 1. This is because the 1JFF-A manifold is a sub-manifold of 1JFF, and all the points of 1JFF-A are subsumed by 1JFF. A discussion about *SVs* between 1JFF, 1JFF-A and 1JFF-B which are obtained in Table I, is given in the Results and Discussion section.

Instead of using the *CV* measure, we define a solid measure by applying the RMSD concept to the structure factor distances using the following procedure. A structure factor is a complex number, so we can embed it as a vector in a 2D Euclidean space. Thus, for each protein, we can define a space with the dimension equal two times the number of computed structure factors. For example, for $l_{max} = 17$ we will have 7770 structure factors, so our space's dimensionality is $2 \times 7770 = 15540 -$ dimension (we represent this space by a 2-column and a 7770-raw matrix). Subsequently, we compute the distances between each pair of elements in this matrix. We obtain an $n \times n$ matrix of the distances. In a similar way we obtain another matrix for the second protein. The next step is to compute the parameter

$$D^2 = 2 \sum_{i<j}^{n} \sum_{j=2}^{n} (d_{ij} - d'_{ij})^2 = 2 \sum_{i<j}^{n} \sum_{j=2}^{n} (d_{ij}^2 + d'^2_{ij} - 2\, d_{ij} d'_{ij})$$

(11)

where $d_{ij}$ and $d'_{ij}$ are the elements of the distance matrix of each of the two proteins. This is an RMSD relation, except we have eliminated the average coefficient $1/(n(n-1))$. Equation 11 in the vector form is

$$D^2 = d^2 + d'^2 - 2\langle \mathbf{d} | \mathbf{d}' \rangle$$

(12)

Where $d^2 = \langle \mathbf{d} | \mathbf{d} \rangle = 2 \sum_{i<j}^{n} \sum_{j=1}^{n} d_{ij}^2$, $d'^2$ is defined similarly (these are vector lengths, i.e., the sum of the squares of arrays), and $\langle \mathbf{d} | \mathbf{d}' \rangle$ is the scalar product of the two protein vectors (i.e., the sum of the corresponding array multiplications). This scalar product indicates the correlation between two proteins, because if there is no correlation, then $D^2 = d^2 + d'^2$, and if we have a maximum correlation (the two proteins are the same), then $D^2 = 0$. To obtain a direct measure of the similarity between two proteins, we define $SV$ by rewriting Eq. 12 as follows:

$$\text{Simalrity Value} = SV = \frac{\langle \mathbf{d} | \mathbf{d}' \rangle}{d^2 + d'^2} = \frac{1}{2}\left(1 - \frac{D^2}{d^2 + d'^2}\right). \tag{13}$$

Based on the above discussion, $SV$ will satisfy the following inequality:

$$\text{not correlated} \leftarrow 0 \leq SV \leq \frac{1}{2} \rightarrow \text{totally correlated.} \tag{14}$$

In other words, when the two proteins are the same, then $SV = 1/2$, and when they are completely different and there is no correlation between them, then $SV = 0$.

The Fourier transformation is a linear transform[21], and it preserves the length and the inner product. Thus, the Fourier transform is an isometric mapping[19,22,23]. We have shown earlier in this paper that the $C_{lmn}$ coefficients are the Fourier transforms of $f$. Therefore, $SV$ is a good measure to compare two proteins. In the following box, we summarize the algorithm for computing the similarity between two proteins in several simple steps.

**Algorithm**

1. Obtain protein data from the PDB website ($xyz$ position coordinates of all atoms).

2. Sort atoms by their distances to the center of mass. It is assumed that all masses are distributed equally for all atoms. Another possibility could be to consider real masses of atoms.

3. Convert Cartesian $xyz$ coordinates of all atoms to the corresponding Euler angles, $\alpha\beta\gamma$ relative to the center of mass of the protein.

4. Define the shape function, $f_i = f(\beta_i, \alpha_i, \gamma_i) = 1, (i = 1,2, \ldots, N)$, $N$: the number of atoms $(\beta_i, \alpha_i, \gamma_i)$ corresponding to the $(x_i, y_i, z_i)$ position coordinates of the $i$th-atom in Euler angles.

5. Compute the results of $D^l_{mn}(\beta_i, \alpha_i, \gamma_i)$ Eqs. 2–4.

6. Compute the structure factor, $C_{lmn}$ from the discrete form of Eq. 8:

$$C_{lmn} = \frac{2l+1}{8\pi^2} \sum_{i=1}^{N} f(\beta_i, \alpha_i, \gamma_i) \ D^{l\ *}_{mn}(\beta_i, \alpha_i, \gamma_i) \ \sin\beta_i \ \Delta\beta_i \ \Delta\alpha_i \ \Delta\gamma_i$$

We sum only over occupied atom positions because the shape function is zero when there is no atom in the voxel. Thus, the other terms are zero.

7. Repeat steps 1–7 for each protein analyzed.

8. Compute $SV$ using Eq. 13 between two proteins selected for comparison.

**Results and Discussion**

Table I lists *CV* and *SV* measures for selected pairs of protein structures as examples.

We see that the correlation value, *CV*, does not give a good comparison between two proteins. This is because it is a criterion to compute the overlap between two manifolds in the reciprocal space. If the two proteins are similar, this criterion gives a good correlation between them because these two proteins have the same structure factors. However, for two different or partially different proteins the *CV* is not very accurate.

To check our *SV* criterion, we have calculated the atomic shape function for the 1JFF-A structure by using the structure factors, $C_{lmn}$. Figure 1 shows the histograms and plots of $f$ and its reconstructions $f_{\text{reconst.}} = \frac{1}{N}\sum_{l_{\max}=1}^{N} f_{l_{\max}}$ for $N = 9$ and $N = 17$ for 1JFF-A. We see that the reconstructed functions, $f_{\text{reconst.}}$, are in good agreement with $f$, especially when $l_{\max}$ increases.

Figure 2 shows that when the surface under a pocket of the structure factors is normalized to one, the structure factor for a given $l$ has the Poissonian distribution:

$$\frac{1}{(2l+1)^2} \frac{\text{abs}(C_{lmn})}{\sqrt{\sum_{m=-l}^{l}\sum_{n=-l}^{l} \text{abs}(C_{lmn})^2}} \rightarrow P_L(t) = \frac{\mu^L}{L!} e^{-\mu} \tag{15}$$

where $\mu = Var(|C_{lmn}|)$ and $L = (2l+1)^2$. The Poissonian distribution is usually considered to be a continuous distribution. However, here we make it discrete since we need to perform a numerical computation. The maximum probability value for the Poissonian distribution occurs when $\mu = L$ and the magnitude of the corresponding peak for probability is then equal to

$$P_L(\mu = L)_{\max} = \frac{L^L}{L!} e^{-L} = \frac{1}{\sqrt{\pi(2L+\frac{1}{3})}} \tag{16}$$

where we used Stirling's approximation relation, i.e. $L! = \sqrt{\pi(2L+\frac{1}{3})}\, L^L e^{-L}$. The peaks in Fig. 2 are in good agreement with Eq. 11. This is another test to confirm the validity of our method, since it gives the same result as the one obtained in x-ray pattern intensity distributions and in Poisson's distribution for random interactions between radiation and matter[24,25].

In the following discussion we wish to highlight the differences between our method and other methods. The methods introduced for comparing protein similarities are normally based on the following the proteome-scale protein structure modeling, score function comparison, obtaining moments or descriptors, comparing RMSD between residues or chains of two proteins[7,26–49]. Discussing all the methods is out of scope in this paper, but here, we review some methods, which may appear similar to our method. One of these methods involves spherical polar Fourier shape density functions (SPF)[26]. This method uses the expansion of the 3D density function in terms of radial and spherical harmonic functions and computes the correlation coefficients between two density function expansion coefficients. The other method uses Zernike descriptors. The Zernike functions are extensions of the spherical harmonic functions. The Zernike descriptors were first used by Novotni and Klein[27] to compare two shapes in shape searching algorithms in computer science. Later they were adapted for protein comparison purposes[28,29]. The 3D Zernike method is a rotational invariant method and it finds a descriptor which represents a given shape. By comparing the descriptors, the similarities between any two shapes could be determined. Another method that should be mentioned here is the spherical harmonic method[7,28]. This method expands a shape function in terms of spherical harmonic functions. After some algebraic computations, the spherical harmonic method defines the descriptors and compares them. The above methods use moment or descriptor concepts to compare proteins. Some methods have used RMSD values as a score to compare between two structures but because of the different protein sizes normally these methods use RMSD only partially. For example, some of these methods have used a difference between the intra-structural residue–residue distances, e.g. Dali[45,46], CE[47], or between inter-structural residue-residue distances, such as STRUCTAL[15], SAL[48] and TM-score[40]. Our method is different from these structural methods for the following reasons:

1. Using the Wigner-D function does not require a definition of a radial function, as is done in the 3D Zernike, spherical harmonic or SPE methods. The three angles in the Wigner-D functions are the Euler angles and it is well known from classical mechanics that moving through a 3D rigid body is possible by using three Euler angles.

2. We show that the expansion coefficients of the shape function defined by the Wigner-D functions are equivalent to the Fourier transform of the shape function (see more details in the Basic Mathematical Idea section). Thus, we introduce the expansion coefficients of a shape function in terms of Wigner-D functions as structure factors.

3. It is well known that the Fourier transform (consequently an expansion on the Wigner-D function) is a linear transform and hence it preserves isometry[17,18,22,23]. Thus, if we define a RMSD-type criterion, like Similarity Value between structure factors, we prove mathematically that the properties obtained in the reciprocal space reflect the same properties in the position space. That means that if in the reciprocal space two proteins are similar, the same result holds true in the position space (provided a method of comparison is defined).

4. The size of a protein analyzed does not affect our comparison. This is because we are able to compare two proteins with the same size in reciprocal space, even though they may be different in position space, and we can choose the dimension of reciprocal space according to a desired level of accuracy. Note that in reality the expansion terms have to go to infinity but similar to other computational calculations, we should choose a cutoff in order to terminate this divergence. The number of expansion coefficients used increases the level of accuracy but also the cost of computation.

5. Contrary to inductive methods, our method is deductive and it is proved by mathematics, thus it does not rely on a great deal of experience to be validated. This is why we did not need to compare a large number of proteins in our manuscript and only a few number of known proteins are given as examples. Nonetheless, we compared our method results with two other sets i.e. 48 set and 86 set, where both liganded, unliganded proteins are listed, and RMSD values in the supplementary material of Li et al.[1] (These sets are reported in http://dragon.bio.purdue.edu/visgrid_suppl). Note that RMSD values are computed in the position space. Tables II and III show that our *SV* values (computed in reciprocal space) are in good agreement with the RMSD values reported in Li et al.[1]. When an RMSD value between two proteins structures increases, *SV* decreases which means that the similarity between these two proteins decreases. Some differences between two results that can be found in the tables are related to the partial computation of RMSD value, especially when the number of atoms in the two proteins compared is different. This good agreement between RMSD values and *SV*s indicates that our *SV* is a reliable parameter for comparing any two proteins. In some algorithms, using RMSD values in a part of the algorithm, RMSD can be replaced by *SV* as an alternative parameter. This is because *SV* is equivalent to RMSD and *SV* can be computed more precisely than RMSD for proteins with different sizes. However, it should be kept in mind that the cost of computation for *SV* is about 33 seconds for comparing two proteins using a laptop with an Intel Core i7 CPU.

Below we discuss a specific example by explaining the *SV*s results for 1JFF and their monomers (1JFF-A and 1JFF-B). As we know, 1JFF-A has $r = 3227 -$ atoms, 1JFF-B has $s = 3351 -$ atoms and 1JFF has the sum of both monomers atoms i.e. $r + s = 6578 -$ atoms. The *SV*s reported between these macromolecules in Table I are:

| 1JFF | 1JFF-A | *SV* = 0.2091 |
| 1JFF | 1JFF-B | *SV* = 0.4967 |
| 1JFF-A | 1JFF-B | *SV* = 0.2271 |

Here, we discuss these results in more detail. To simplify notation, we represent 1JFF-A with A, 1JFF-B with B, and 1JFF with AB. Their distance matrices are defined by **d**(A,A) which has $r \times r$ arrays, similarly **d**(B,B) with $s \times s$ matrix elements, and **d**(AB,AB) with $(r+s) \times (r+s)$ arrays. Note that here we define distance matrix **d** in position space. We can write **d**(AB,AB) as

$$\mathbf{d}(AB,AB) = \mathbf{d}(A,A) \oplus \mathbf{d}(A,B) \oplus \mathbf{d}(B,A) \oplus \mathbf{d}(B,B) \quad (17)$$

where $\oplus$ indicates direct sum between matrices (see Fig. 3). We note that **d**(A,B) and **d**(B,A) are transpose of each other. Let us assume an unknown direct way (in position space and not in reciprocal space), then we can find the RMSD between the above structures. We then compute the following terms

$$\begin{aligned} D^2(A,B) &= \sum [\mathbf{d}(A,A) \ominus \mathbf{d}(B,B)].^\wedge 2 \\ D^2(AB,A) &= \sum [\mathbf{d}(AB,AB) \ominus \mathbf{d}(A,A)].^\wedge 2 \\ D^2(AB,B) &= \sum [\mathbf{d}(AB,AB) \ominus \mathbf{d}(B,B)].^\wedge 2 \end{aligned} \quad (18)$$

where $[\cdot].^\wedge 2$ means that all arrays of the matrix in the bracket will be squared, $\sum [\cdot]$ is defined as summation over all matrix arrays in the bracket, $[\cdot]$, $\ominus$ indicates "an imaginary minus sign" between two different size matrices (the difference between matrix sizes is a serious problem when trying to define a corresponding RMSD) and we do not know how it acts. Now, we expand $D^2$ between AB and A and B. First, we have

$$D^2(\text{AB,A}) = \sum [\mathbf{d}(\text{AB,AB}) \ominus \mathbf{d}(\text{A,A})]^2$$

$$= \sum [\mathbf{d}(\text{A,A}) \oplus \mathbf{d}(\text{A,B}) \oplus \mathbf{d}(\text{B,A}) \oplus \mathbf{d}(\text{B,B}) \ominus \mathbf{d}(\text{A,A})].\text{\textasciicircum}2$$

(19)

Similarly, we find that

$$D^2(\text{AB,B}) = \sum [\mathbf{d}(\text{AB,AB}) \ominus \mathbf{d}(\text{B,B})].\text{\textasciicircum}2$$

$$= \sum [\mathbf{d}(\text{A,A}) \oplus \mathbf{d}(\text{A,B}) \oplus \mathbf{d}(\text{B,A}) \oplus \mathbf{d}(\text{B,B}) \ominus \mathbf{d}(\text{B,B})].\text{\textasciicircum}2$$

(20)

We readily see that $D^2(\text{AB, A}) \neq D^2(\text{AB, B})$. This shows that two monomers do not have the same RMSD values or *SV*s.

Now, one can add leaks of arrays in distance matrix of A (or B) with respect to the distance matrix size of AB by adding zeros (see Fig. 4). Note that this assumption has not been proved and is only a heuristic. Thus, the following relation could be obtained (a derivation is given by Figs. 5 and 6)

$$|D^2(\text{AB, A}) - D^2(\text{AB, B})| = |d^2(\text{B, B}) - d^2(\text{A, A})| \qquad (21)$$

where $d^2(\text{B, B}) = \langle \mathbf{d}(\text{B, B}) | \mathbf{d}(\text{B, B}) \rangle$ has a similar definition to the one mentioned after Eq. 12. However, this definition does not provide a normalized measure to compare with the *SV*. Now, we define *SS* as follows

$$SS = |S(\text{AB, A}) - S(\text{AB, B})| \cong \left| \frac{d^2(\text{B,B}) - d^2(\text{A,A})}{2\left(d^2(\text{AB,AB}) + d^2(\text{A,A})\right)} \right| \qquad (22)$$

where we have defined $S(\text{AB, A})$ as follows

$$S(\text{AB, A}) = \frac{D^2(\text{AB,A})}{2\left(d^2(\text{AB,AB}) + d^2(\text{A,A})\right)} \qquad (23)$$

and $S(AB, B)$ is defined similarly. Here, to arrive at the right-hand side of Eq. 22 we approximated $d^2(A, A) \cong d^2(B, B)$ in the dominator. We will see later this is a very reasonable approximation.

The computation of $d^2$ for 1JFF and its two monomers 1JFF-A and 1JFF-B results in $d^2(AB, AB) = 7.1519 \times 10^{10}$, $d^2(A, A) = 8.7874 \times 10^9$ and $d^2(B, B) = 9.7127 \times 10^9$ in units of square Angstroms. From these values, it is easy to obtain $d^2(A, B) = d^2(B, A) = 2.6509 \times 10^{10}$. To get a sense of the numerical values involved, in addition to our real case, we compute $SS$ in Eq. 22 for a totally correlated ($d^2(AB, AB) = 0$) case and for an uncorrelated ($d^2(AB, AB) = d^2(B, B) + d^2(A, A)$) case. Thus, we have

$$SS \cong \begin{cases} \frac{1}{6}\left(1 - \frac{d^2(A,A)}{d^2(B,B)}\right) = 0.0159 & \text{A and B are totally different } d^2(A,B) = 0 \\ 0 & \text{A and B are totally similar} \\ 0.0058 & \text{In our case} \end{cases} \quad (24)$$

The above equation shows that $0 \leq SS \leq 0.0159$. We normalize $SS$ to ½ so that we have

$$SS_{\text{Normalized}} = \frac{1}{2}\left(1 - \frac{SS}{0.0159}\right) = \begin{cases} 0 & \text{A and B are totally different} \\ \frac{1}{2} & \text{A and B are totally similar} \end{cases} \quad (25)$$

Thus, for our case we have: $SS_{\text{Normalized}} = 0.3176$. Now, we come back to $SV$s obtained in Table I that yields: $SV(AB, A) - SV(AB, B) = 0.4967 - 0.2091 = 0.2876$. We see that these two results are in good agreement. Note that $SS_{\text{Normalized}}$ is obtained by using an approximation. Here in the structural form, we have shown why two monomers of 1JFF are different. Thus, if one of these monomers is similar to 1JFF the other could not be similar and vice versa.

**Conclusions**

This paper introduces a new method to compare protein structures; it can be generalized to compare arbitrary shapes defined as a set of 3D coordinates. The novelty of our method lies in expanding the shape function using Wigner-D functions, showing that the expansion coefficients correspond to the structure factors, and using the RMSD measure in the reciprocal space (for the structure factors) to define a similarity value, namely the $SV$ parameter. We show that this measure gives a corresponding similarity in the spatial domain because of the isometric property of the Fourier transform. We have verified our method by obtaining the shape function by using the structure factors and Wigner-D functions (see Fig. 2). The absolute values of the structure factors are the same as the intensities measured by x-ray scattering. We also show that the structure factor distribution is a Poisson distribution; as is well known, the intensity distribution for x-ray scattering is also a Poisson distribution. This result demonstrates the reliability of our method. The numerical results shown in Tables I–III for *SV* also confirm the reliability and usefulness of our method.

An important problem for similarity comparison methods is that the number of the protein atoms in an arbitrary pair of proteins is generally not the same. To address this problem, some methods use partial similarity measures between two proteins. However, in our method, despite the fact that the number of atoms of the two proteins being compared is different, the number of structure factors is the same in reciprocal space. This is another important advantage of our method.

**Acknowledgements**

S.M.S.F. acknowledge grant number 2/22306 from Ferdowsi University of Mashhad. J.A.T. gratefully acknowledges research support received from the National Science and Engineering Research Council (Canada), the Canadian Breast Cancer Foundation, and the Allard Foundation.

**Figure Legends**

**Fig. 1.** Histograms (a) and plots (b) of the shape function $f$ (blue) and its reconstructions $f_{reconst.}$ for $N = l_{max} = 9$ (green) and $N = l_{max} = 17$ (red) for 1JFF-A. We see that the reconstructed $f_{reconst.}$ are in good agreement with $f$ when $l_{max}$ increases.

**Fig. 2.** Panel (a) shows non-normalized absolute values of structure factors. The abscissa is the total number of structure factors. Panel (b) shows the normalized area under the curve to one of the absolute values of the structure factors for each $l$. The abscissa is the value of $(2l + 1)^2$.

**Fig. 3.** A schematic of the distance matrix representation.

**Fig. 4.** Extended $\mathbf{d}(B, B)$ matrix. Here we add zeros to change the size of $\mathbf{d}(B, B)$ from $s \times s$ to $(r + s) \times (r + s)$.

**Fig. 5.** A schematic representation of obtaining $\mathbf{d}(AB, B)$. The subtraction of the two top matrices yields the bottom matrix.

**Fig. 6. Top.** Squared arrays of $\mathbf{d}(AB, A)$ and $\mathbf{d}(AB, B)$. ".^2" means that all arrays of the matrix will be squared. The sum over all arrays of these matrices yields $D^2$ as defined by Eq. 11. **Bottom.** Subtraction of two top matrices. Normally, we should subtract the sum of two top matrices. But, here to show the derivation of our formula in Eq. 21 before summation we subtract two top matrices and we see the result in the bottom. To solve Eq. 21, we have to sum over all arrays of bottom matrix. The minus sign causes the change of shading on $\mathbf{d}(A, A).^2$ in the bottom matrix.

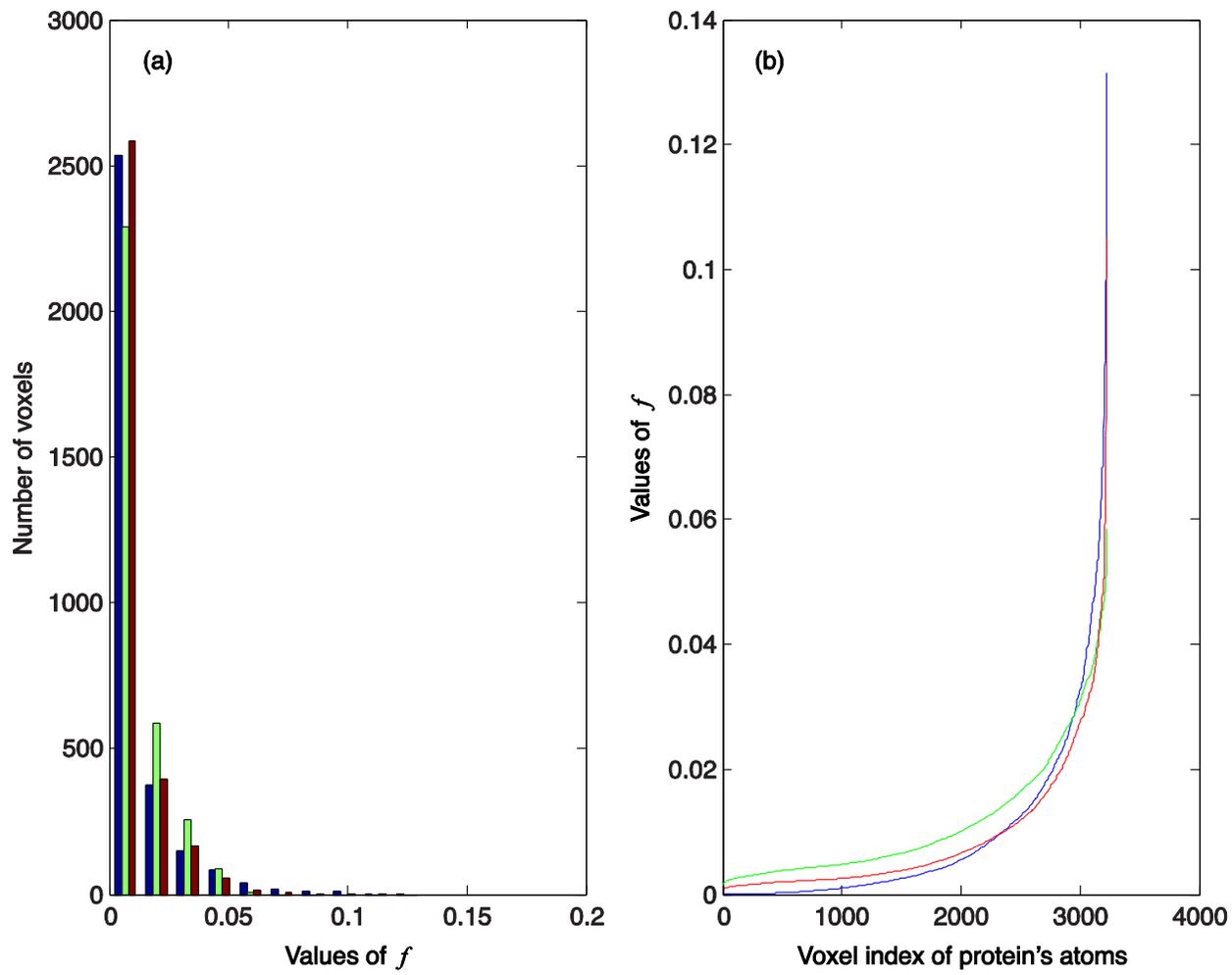

**Fig. 1**

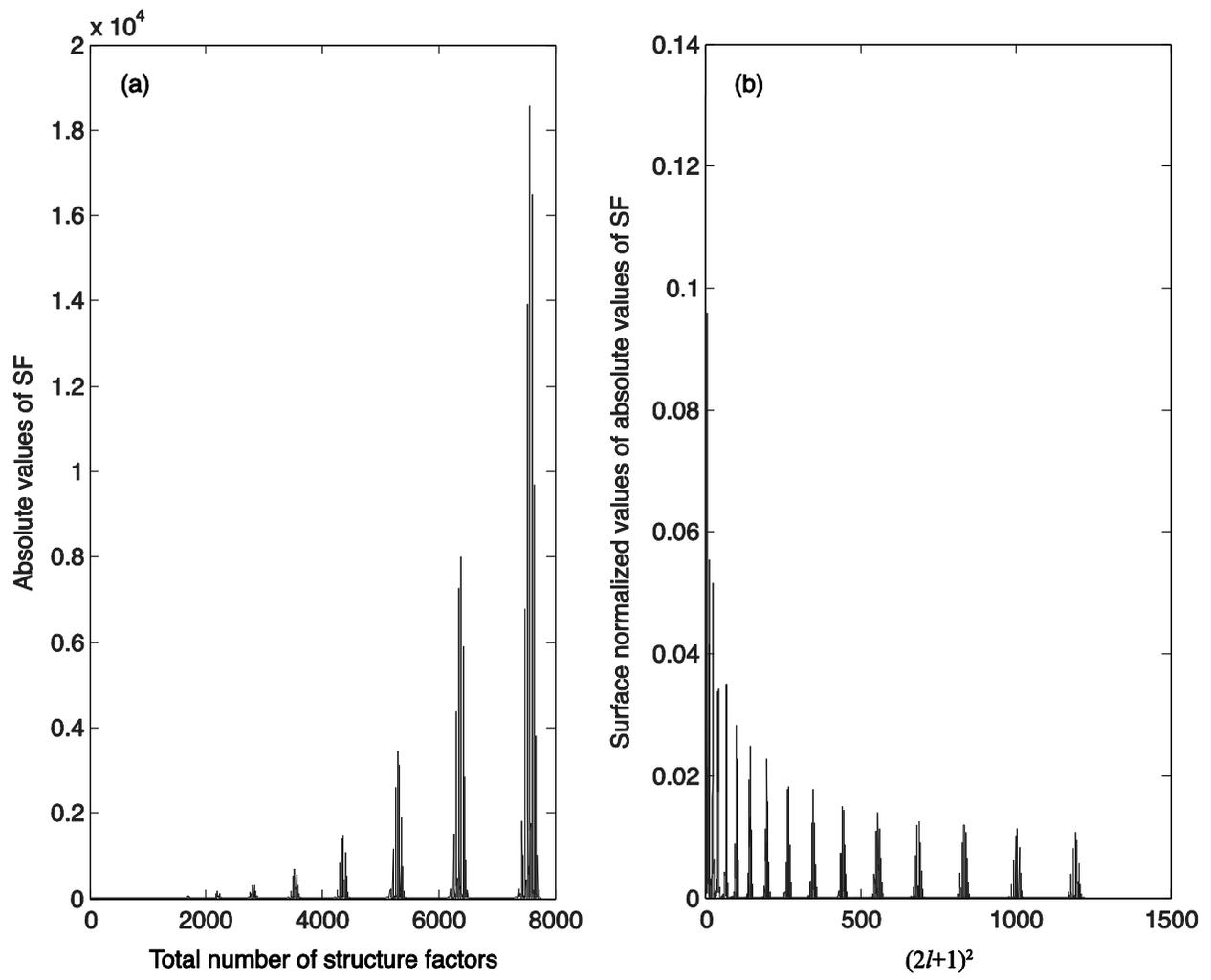

**Fig. 2**

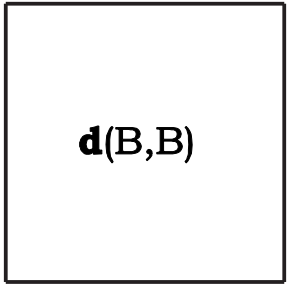 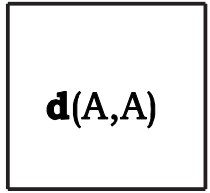 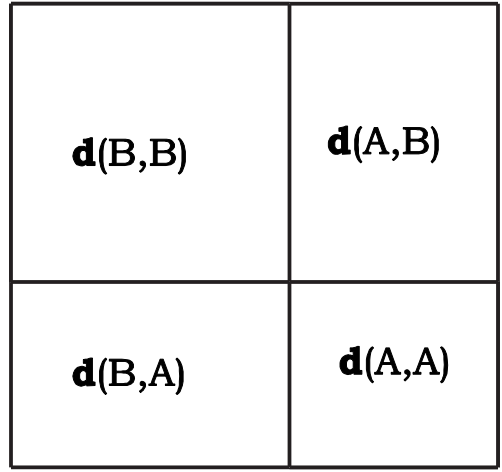

**Fig. 3**

## Extended **d**(B,B) matrix

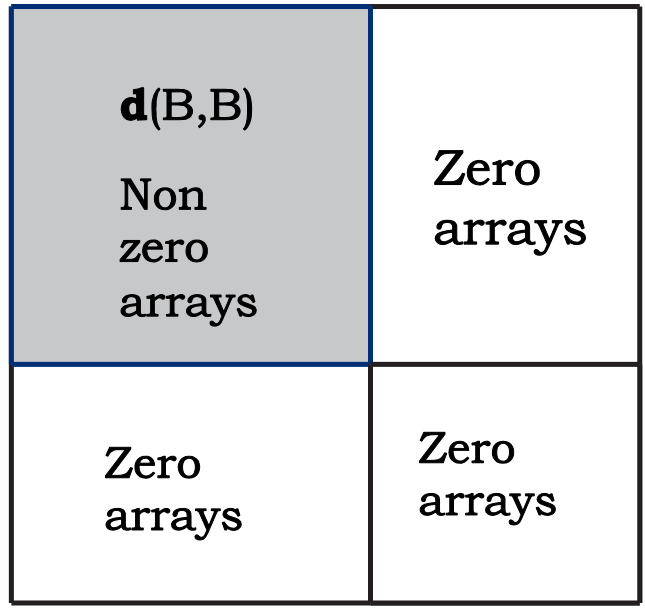

**Fig. 4**

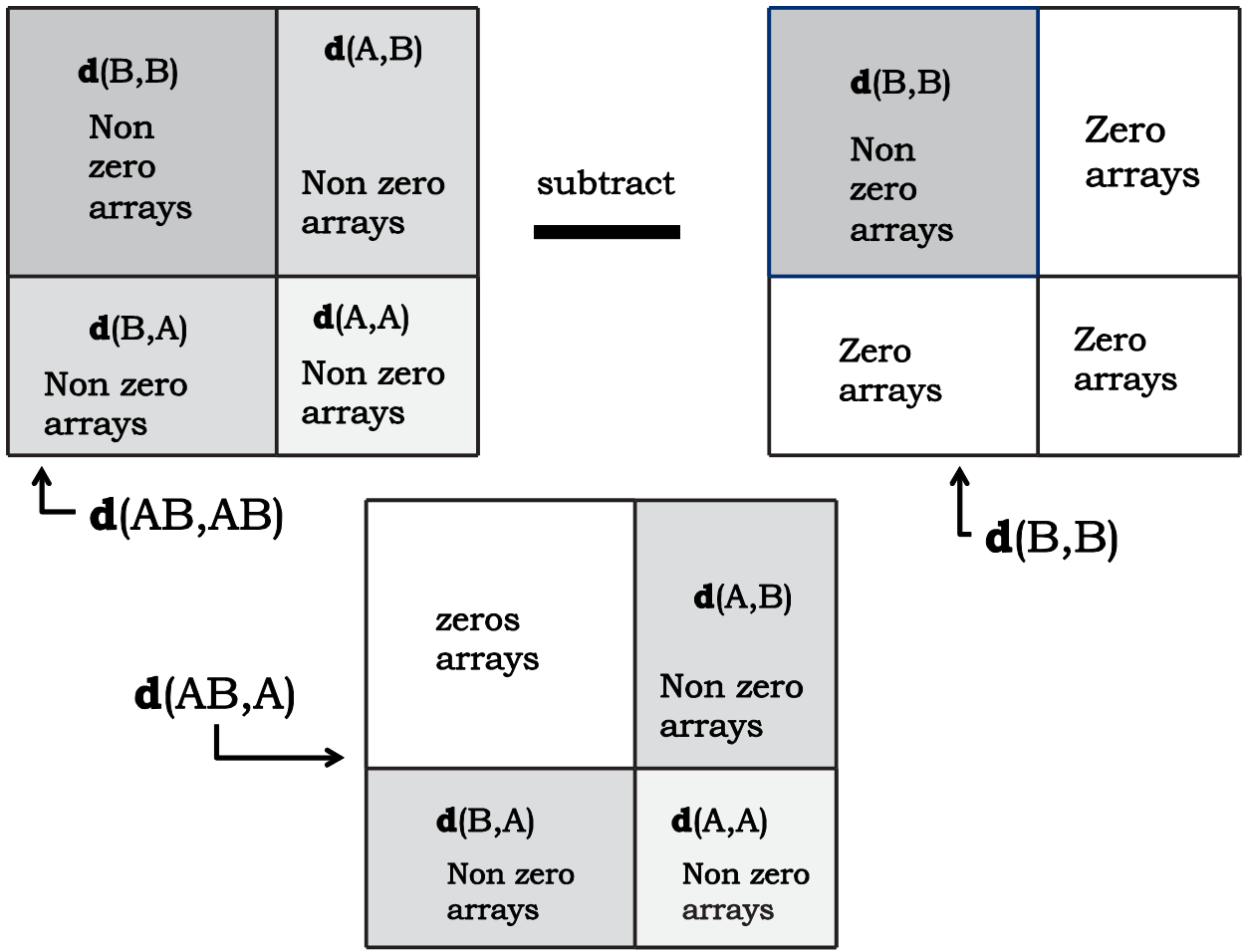

Fig. 5

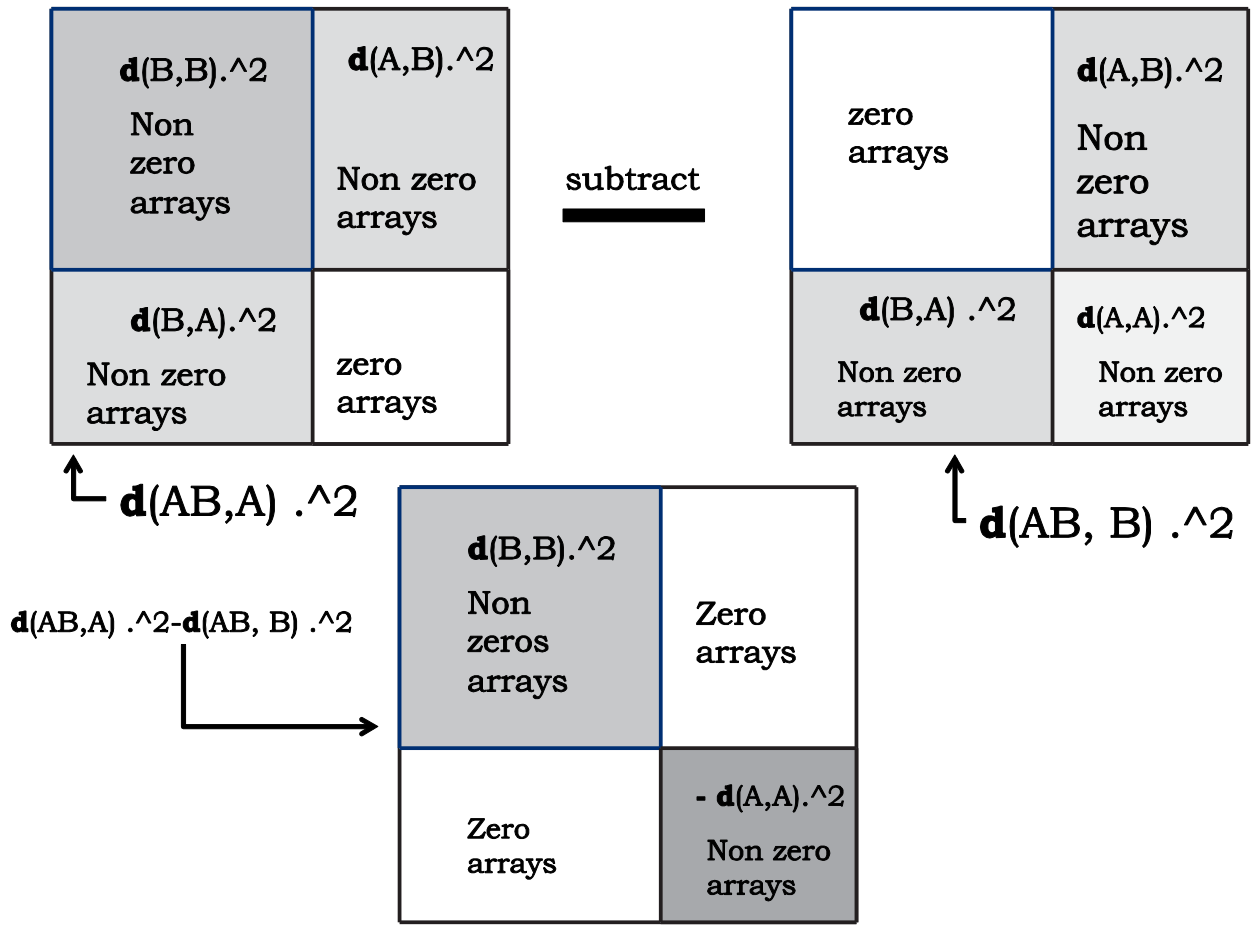

Fig. 6

**Table I.** Comparison between pairs of proteins using correlation value (*CV*) and similarity value (*SV*). 1JFF, 1SA0, 1TUB, and 1FSZ are all structures of proteins in the tubulin-FtsZ superfamily. 1ATN is a structure for actin. Comparisons between unrelated protein pairs (tubulin-FtsZ superfamily with actin) are italicized.

| First Protein's PDB ID | Second Protein's PDB ID | Correlation Value (*CV*) | Similarity Value (*SV*) |
|---|---|---|---|
| *1JFF* | *1ATN* | *0.3962* | *0.0002* |
| 1JFF | 1FSZ | 0.9661 | 0.3453 |
| 1JFF | 1SA0 | 0.9361 | 0.0981 |
| 1JFF | 1TUB | 0.9933 | 0.4872 |
| 1JFF | 1JFF-A | 0.9537 | 0.2091 |
| 1JFF | 1JFF-B | 0.9987 | 0.4967 |
| 1JFF-A | 1JFF-B | 0.9460 | 0.2271 |
| *1JFF-A* | *1ATN* | *0.5634* | *0.0010* |
| 1JFF-A | 1SA0 | 0.9955 | 0.3771 |
| 1JFF-A | 1TUB | 0.9300 | 0.2425 |
| 1JFF-A | 1FSZ | 0.9795 | 0.4131 |
| 1JFF-B | 1SA0 | 0.9270 | 0.1073 |
| *1JFF-B* | *1ATN* | *0.3782* | *0.0002* |
| 1JFF-B | 1TUB | 0.9969 | 0.4962 |
| 1JFF-B | 1FSZ | 0.9637 | 0.3689 |
| *1ATN* | *1SA0* | *0.5973* | *0.0023* |
| *1ATN* | *1FSZ* | *0.5016* | *0.0005* |
| *1ATN* | *1TUB* | *0.3639* | *0.0002* |
| 1SA0 | 1FSZ | 0.9806 | 0.2322 |
| 1SA0 | 1TUB | 0.9085 | 0.1154 |
| 1FSZ | 1TUB | 0.9489 | 0.3852 |

**Table II.** Set of 48 protein structures with *SV* and RMSD from Li et al.[1] for comparison. The *SV*s are computed from structure factors for $l_{max} = 7$.

| First Protein's PDB ID | Second Protein's PDB ID | Similarity Value (*SV*) | RMSD [1] |
|---|---|---|---|
| 1A6W | 1A6U | 0.198 | 0.34 |
| 1MRG | 1AHC | 0.401 | 0.43 |
| 1RNE | 1BBS | 0.316 | 0.61 |
| 1RBP | 1BRQ | 0.108 | 0.62 |
| 1BYB | 1BYA | 0.499 | 0.43 |
| 1HFC | 1CGE | 0.399 | 0.37 |
| 3GCH | 1CHG | 0.070 | 1.10 |
| 1BLH | 1DJB | 0.497 | 0.23 |
| 1INC | 1ESA | 0.397 | 0.21 |
| 1GCA | 1GCG | 0.499 | 0.32 |
| 1HEW | 1HEL | 0.498 | 0.21 |
| 1IDA | 1HSI | 0.083 | 1.07 |
| 1DWD | 1HXF | 0.150 | 0.27 |
| 2IFB | 1IFB | 0.382 | 0.37 |
| 1IMB | 1IME | 0.498 | 0.22 |
| 2PK4 | 1KRN | 0.445 | 0.39 |
| 2TMN | 1L3F | 0.266 | 0.62 |
| 1IVD | 1NNA | 0.426 | 1.23 |
| 1HYT | 1NPC | 0.332 | 0.88 |
| 1PDZ | 1PDY | 0.499 | 0.66 |
| 1PHD | 1PHC | 0.499 | 0.17 |
| 1PSO | 1PSN | 0.499 | 0.33 |
| 1SRF | 1PTS | 0.498 | 0.26 |
| 1ACJ | 1QIF | 0.497 | 0.31 |
| 1SNC | 1STN | 0.495 | 0.70 |
| 1STP | 1SWB | 0.145 | 0.33 |
| 1ULB | 1ULA | 0.474 | 0.79 |
| 2YPI | 1YPI | 0.165 | 1.27 |
| 2H4N | 2CBA | 0.498 | 0.20 |
| 2CTC | 2CTB | 0.499 | 0.15 |
| 5CNA | 2CTV | 0.034 | 0.40 |
| 1FBP | 2FBP | 0.494 | 1.06 |
| 2SIM | 2SIL | 0.499 | 0.14 |
| 1MTW | 2TGA | 0.159 | 0.42 |
| 1APU | 3APP | 0.498 | 0.40 |
| 1QPE | 3LCK | 0.465 | 0.28 |
| 5P2P | 3P2P | 0.480 | 0.42 |
| 4PHV | 3PHV | 0.045 | 1.23 |
| 3PTB | 3PTN | 0.122 | 0.26 |
| 1BID | 3TMS | 0.499 | 0.24 |
| 1OKM | 4CA2 | 0.472 | 0.22 |
| 4DFR | 5DFR | 0.496 | 0.82 |
| 3MTH | 6INS | 0.381 | 1.09 |
| 6RSA | 7RAT | 0.440 | 0.18 |
| 1CDO | 8ADH | 0.403 | 1.34 |
| 7CPA | 5CPA | 0.132 | 0.40 |
| 1ROB | 8RAT | 0.469 | 0.28 |
| 1IGJ | 1A4J | 0.411 | 0.80 |

**Table III.** Set of 86 protein structures with *SV* and RMSD from Li et al.[1] for comparison. The *SV*s are computed from structure factors for $l_{max} = 7$.

| First Protein's PDB ID | Second Protein's PDB ID | Similarity Value (*SV*) | RMSD [1] |
|---|---|---|---|
| 1AD4 | 1AD1 | 0.499 | 0.50 |
| 1AHX | 1AHG | 0.499 | 0.24 |
| 1AUR | 1AUO | 0.499 | 0.20 |
| 1AXZ | 1AXY | 0.498 | 0.12 |
| 1GN8 | 1B6T | 0.491 | 0.51 |
| 1B9Z | 1B90 | 0.494 | 0.54 |
| 1LRI | 1BEO | 0.498 | 1.05 |
| 1BUL | 1BUE | 0.499 | 0.18 |
| 1BYD | 1BYA | 0.499 | 0.43 |
| 1C3R | 1C3P | 0.202 | 0.39 |
| 1C5I | 1C5H | 0.494 | 0.13 |
| 1QJW | 1CB2 | 0.498 | 0.63 |
| 1CTE | 1CPJ | 0.499 | 0.29 |
| 1SZJ | 1CRW | 0.499 | 0.33 |
| 1ESW | 1CWY | 0.498 | 0.38 |
| 1CY7 | 1CY0 | 0.155 | 1.12 |
| 1DED | 1D7F | 0.481 | 0.26 |
| 1P7T | 1D8C | 0.406 | 0.66 |
| 1DMY | 1DMX | 0.499 | 0.19 |
| 1DQY | 1DQZ | 0.052 | 0.75 |
| 1LP6 | 1DV7 | 0.471 | 0.56 |
| 1E2S | 1E1Z | 0.499 | 0.13 |
| 1ESE | 1ESC | 0.499 | 0.19 |
| 6ALD | 1EWD | 0.477 | 0.44 |
| 1NLM | 1F0K | 0.163 | 1.66 |
| 1F4X | 1F4W | 0.488 | 0.25 |
| 1JBW | 1FGS | 0.430 | 1.48 |
| 1FR8 | 1FGX | 0.498 | 0.54 |
| 1LD8 | 1FT1 | 0.416 | 0.92 |
| 1HVC | 1G6L | 0.345 | 0.46 |
| 1LSP | 1GBS | 0.360 | 0.26 |
| 1LC3 | 1GCU | 0.458 | 0.77 |
| 1GJW | 1GJU | 0.499 | 0.29 |
| 1N75 | 1GLN | 0.422 | 1.47 |
| 1GOY | 1GOU | 0.476 | 0.47 |
| 1H46 | 1GPI | 0.193 | 0.15 |
| 1GUZ | 1GV1 | 0.383 | 0.62 |
| 1YDD | 1HEA | 0.491 | 0.18 |
| 1YDA | 1HEB | 0.498 | 0.20 |
| 1KIC | 1HOZ | 0.420 | 0.35 |
| 1A80 | 1HW6 | 0.466 | 0.93 |
| 1I3A | 1I39 | 0.498 | 0.40 |
| 4AIG | 1IAG | 0.494 | 0.26 |
| 1JZS | 1ILE | 0.497 | 0.69 |
| 1JQ3 | 1INL | 0.493 | 0.35 |
| 1JAY | 1JAX | 0.435 | 0.60 |
| 1UEH | 1JP3 | 0.499 | 0.67 |
| 1JSO | 1JSM | 0.499 | 0.10 |
| 1JYL | 1JYK | 0.208 | 0.94 |
| 1JVS | 1K5H | 0.351 | 1.16 |
| 1K70 | 1K6W | 0.497 | 1.08 |
| 1M6P | 1KEO | 0.136 | 1.05 |
| 3KIV | 1KIV | 0.467 | 0.30 |
| 1KMP | 1KMO | 0.498 | 0.64 |
| 2NGR | 1KZ7 | 0.467 | 1.61 |
| 2MIN | 1L5H | 0.084 | 0.55 |
| 1LL2 | 1LL3 | 0.496 | 0.37 |
| 1LMC | 1LMN | 0.499 | 0.10 |
| 1EYN | 1NAW | 0.208 | 1.02 |
| 1BHT | 1NK1 | 0.033 | 0.58 |
| 1PBO | 1OBP | 0.143 | 0.38 |
| 1OPB | 1OPA | 0.295 | 0.68 |
| 1I75 | 1PAM | 0.499 | 0.13 |
| 1NME | 1PAU | 0.499 | 0.29 |
| 1KEV | 1PED | 0.281 | 0.81 |
| 1PIG | 1PIF | 0.495 | 0.32 |
| 1PJC | 1PJB | 0.498 | 0.61 |
| 1KLT | 1PJP | 0.168 | 0.97 |
| 1QHG | 1PJR | 0.499 | 0.23 |
| 1CEB | 1PKR | 0.041 | 0.58 |
| 2PK4 | 1PMK | 0.417 | 0.71 |
| 1BK9 | 1PSJ | 0.494 | 0.24 |
| 1QBB | 1QBA | 0.497 | 0.11 |
| 1PYY | 1QME | 0.157 | 0.59 |
| 1OSS | 1SGT | 0.367 | 0.27 |
| 1SWN | 1SWL | 0.497 | 0.31 |
| 1LBT | 1TCA | 0.440 | 0.24 |
| 1WBL | 1WBF | 0.371 | 0.39 |
| 1YDB | 1YDC | 0.491 | 0.12 |
| 1H0S | 2DHQ | 0.499 | 0.26 |
| 1LLO | 2HVM | 0.498 | 0.12 |
| 43CA | 43C9 | 0.491 | 0.23 |
| 5BIR | 4BIR | 0.487 | 0.61 |
| 5EUG | 4EUG | 0.498 | 0.21 |
| 5EAU | 5EAS | 0.064 | 0.40 |
| 7TAA | 6TAA | 0.499 | 0.24 |